\documentclass[aps,twocolumn,noshowkeys]{revtex4-1}
\usepackage{amsmath}
\usepackage{graphicx}
\usepackage{bm}
\usepackage{hyperref}
\usepackage{xcolor}
\usepackage{multirow}

\newcommand{\Eref}[1]{Eq.~(\ref{#1})}
\newcommand{\tref}[1]{Table~\ref{#1}}

\newcommand{\WaxeN}{W_{\textrm{ax}}^{(eN)}}
\newcommand{\Waxee}{W_{\textrm{ax}}^{(ee)}}
\newcommand{\angstrom}{\mbox{\normalfont\AA}}

\begin{document}
\title{Axion-mediated electron-electron interaction in ytterbium monohydroxide molecule}

\begin{abstract}
The YbOH triatomic molecule can be efficiently used to measure the electron electric dipole moment, which violates time-reversal ($T$) and spatial parity ($P$) symmetries of fundamental interactions [I.  Kozyryev,  N. R.  Hutzler,  Phys.  Rev.  Lett. \textbf{119}, 133002 (2017)]. We study another mechanism of the $T,P$-violation in the YbOH molecule -- the electron-electron interaction mediated by the low-mass axionlike particle. For this we calculate the molecular constant that characterizes this interaction and use it to estimate the expected magnitude of the effect to be measured. It is shown that this molecular constant has the same order of magnitude as the corresponding molecular constant corresponding to the axion-mediated electron-nucleus interaction. According to our estimation, an experiment on YbOH will allow one to set updated laboratory constraints on the $CP$-violating electron-axion 
coupling
constants.
\end{abstract}

\author{D.E. Maison$^{1,2}$}
\email{maison$\_$de@pnpi.nrcki.ru}
\author{L.V.\ Skripnikov$^{1,2}$}
\author{A.V. Oleynichenko$^{1,3}$}
\author{A.V. Zaitsevskii$^{1,3}$}
\affiliation{$^{1}$Petersburg Nuclear Physics Institute named by B.P.\ Konstantinov of National Research Center ``Kurchatov Institute'' (NRC ``Kurchatov Institute'' - PNPI), 1 Orlova roscha, Gatchina, 188300 Leningrad region, Russia}
\homepage{http://www.qchem.pnpi.spb.ru}
\affiliation{$^{2}$Saint Petersburg State University, 7/9 Universitetskaya nab., St. Petersburg, 199034 Russia}
\affiliation{$^{3}$Department of Chemistry, M.V. Lomonosov Moscow State University, Leninskie gory 1/3, Moscow, 119991~Russia}

\maketitle

\section{Introduction}

One of the most intriguing problems of modern theoretical physics is the development of the unified theory of physical interactions. The development of the standard model of elementary particles (SM) in 1970s was a significant milestone in this direction. However, some observable phenomena cannot be explained within the SM. There is no generally accepted theoretical description of cosmological dark matter and dark energy, which are indicated by astronomical and cosmological observations. It is also unclear why the combined charge ($C$) and spatial parity ($P$) symmetry 
$CP$
seems to be preserved in the quantum chromodynamics (QCD) sector. It is known that the $CP$-violation is a necessary condition for the baryogenesis at the early stage of the Universe evolution \cite{Sakharov1967}.

Various hypothetical dark matter particles have not been yet reliably detected, despite of numerous attempts \cite{Drukier1986, Ahmed2011, Aprile2017,Aalseth2011, adhikari2019experiment, kane2008dark}. One of the candidates for the dark matter particle is the axion or axionlike particle (ALP) -- the hypothetical pseudoscalar particle with nonzero mass. According to Weinberg~\cite{weinberg1978new} and Wilczek~\cite{wilczek1978problem}, axion arises as a quasi-Nambu-Goldstone boson due to the spontaneous breaking of the Peccei-Quinn symmetry $U_{\textrm{PQ}}(1)$~\cite{peccei1977cp}. Originally, this symmetry breaking was suggested as a part of the strong $CP$-problem solution of QCD. Later, it was
also realized that the axion satisfies all the requirements to be the dark matter component \cite{preskill1983cosmology, abbott1983cosmological,dine1983not}. So, observation of the axion or ALP would be an essential step in the solution of two modern physics problems from areas of cosmology and QCD. The constraints on the axion properties obtained in axion-search experiments can be found in, e.g., Refs.
\cite{Hare:2020,youdin1996limits, ni1999search, Duffy:2006, Zavattini:2006, hammond2007new, hoedl2011improved, barth2013cast, pugnat2014search, flambaum2018resonant,aybas2021,Roussy:2021}.

One of the techniques to search for ALPs in laboratory tabletop experiments with atoms and molecules was suggested in Ref.~\cite{graham2013new}.
It is based on measurements of oscillating 
$T,P$-violating
atomic and molecular electric dipole moments (EDM) arising due to the interaction with the oscillating ALP field, where $T$ is the time reversal operation. The frequency of oscillations is determined by the ALP mass and the amplitude is proportional to $\sqrt{\rho}$, where $\rho$ is the ALP density \cite{Graham:2011,graham2013new}. The latter can be estimated using the assumption that the dark matter is dominated by ALPs.
The recent Cosmic Axion Spin Precession Experiment (CASPEr)~\cite{budker2014proposal} was aimed to measure this effect using the solid-state nuclear magnetic resonance in a polarized ferroelectric crystal and some results have been recently reported~\cite{aybas2021}. In this experiment the oscillating ALP field leads to the oscillating nuclear Schiff moment that interacts~\cite{Mukhamedjanov:05,Ludlow:13,Skripnikov:16a} with the crystal environment which can be detected. Molecular constraints have been recently obtained using the HfF$^+$ cation~\cite{Roussy:2021} assuming oscillations of the scalar-pseudoscalar nucleus-electon interaction constant, taking into account experimental data~\cite{Cornell:2017} as well as theoretical values of the molecular constant characterizing the scalar-pseudoscalar interaction from Refs.~\cite{Skripnikov:17c,Fleig:17}.

In the present paper we consider the appearance of a \textit{static} molecular EDM, induced by the virtual ALP exchange between electrons.
It was shown in Ref.~\cite{Stadnik:2018} that the ALP-mediated interaction between fermions in atomic and molecular systems can be one of the sources of $T,P$-violating interactions that can be probed in experiments aimed at searches for the electron electric dipole moment ($e$EDM). 
In Ref. \cite{Stadnik:2018} one can find the interpretation of recent experiments with atoms \cite{murthy1989new, Regan:02, Graner:2016, rosenberry2001atomic} and diatomic molecules \cite{Hudson:02, Cornell:2017, ACME:14a} in terms of the ALP mass and ALP-fermion interaction constants. A feature of such experiments is that they are independent of cosmological assumptions.

It was recently proposed to perform experiments to search for $T,P$-violating effects such as the $e$EDM using the YbOH linear triatomic molecule~\cite{kozyryev2017precision}. Its advantage over similar diatomics consists in the existence of closely lying energy levels of opposite parity due to the $l$-doubling effect \cite{kozyryev2017precision, Hutzler:2020}. Owing to this feature, such molecules can be polarised using weak laboratory electric fields, reducing systematic effects. Besides, YbOH molecules can be cooled to extremely low temperatures \cite{augenbraun2020laser}. Therefore, it is possible to achieve a large coherence time and increase experimental sensitivity significantly. So, it is expected that current experiments with triatomic molecules would give updated constraints on $e$EDM and other parameters of the $T,P$-violating effects. Due to the arising interest in this area, a number of $T,P$-violation effects in the YbOH molecule have been extensively studied theoretically by several groups \cite{denis2019enhancement,prasannaa2019enhanced,gaul2020ab,denis2020enhanced,Maison:2019b}.
In the paper \cite{Maison:2021} the $T,P$-violating ALP-mediated nucleon-electron interaction in YbOH has been considered. 
All of these studies were devoted to the ``one-electron'' interactions, i.e. to interactions described by the operator whose electronic part depends on coordinates of only one electron. In the present paper we consider the $T,P$-violating interaction between \textit{two} electrons that is mediated by the ALP. The treatment of such interactions drastically differs from previous studies~\cite{denis2019enhancement,Maison:2019b,prasannaa2019enhanced,denis2020enhanced,gaul2020ab,Maison:2021} from the electronic structure theory point of view. It is shown that this interaction should not be less important than the electron-nucleus ALP-mediated interaction from the molecular theory side.

\section{Theory}
The Lagrangian of the interaction of ALP with fermions is given by \cite{moody1984new}:
\begin{equation}\label{lagrangian}
    \mathcal{L}_{\textrm{int}} = a \sum_\psi \bar{\psi} \left(g_\psi^s + ig_\psi^p \gamma_5 \right) \psi \ .
\end{equation}
Here $a$ is the axion field, $\psi$ is the fermion field and $\bar{\psi} = \psi^{+} \gamma_0$, $g_\psi^s$ and $g_\psi^p$ denote the scalar and pseudoscalar coupling constants, respectively, and $\gamma$ are Dirac matrices, defined according to Ref. \cite{Khriplovich:91}. The summation in (\ref{lagrangian}) is performed over all the fermion types in the system. In the present paper we investigate the case of $\psi$ being the electron field. 

The interaction of two electrons, mediated by the axion with the mass $m_a$ can lead to the Yukawa-type interaction potential \cite{moody1984new,Stadnik:2018}:
\begin{equation} \label{potential}
    V_{ee}(\boldsymbol{r}_1, \boldsymbol{r}_2) = 
    +i\frac{g_e^s g_e^p}{4\pi} \frac{e^{-m_a |\boldsymbol{r}_1 - \boldsymbol{r}_2}|}{|\boldsymbol{r}_1 - \boldsymbol{r}_2|} \gamma_0 \gamma_5 \ .
\end{equation}
In this equation $\boldsymbol{r}_1$ and $\boldsymbol{r}_2$ are positions of these electrons and $\gamma$ matrices refer to the first electron. The QCD axion models assume that the $g_e^s$, $g_e^p$ and $m_a$ are related to each other \cite{moody1984new,stadnik:18a}. However, there are numerous models of ALPs assuming these parameters to be independent. In the present paper we consider this commonly used generalized definition and do not distinguish axions and ALPs.

The analogous $T,P$-violating axion-mediated nucleon-electron interaction has the following form:
\begin{equation}
    V_{eN} (\boldsymbol{r}) = +i \frac{g_N^sg_e^p}{4 \pi}
    \frac{e^{-m_a |\boldsymbol{r} - \boldsymbol{R}|}}{|\boldsymbol{r} - \boldsymbol{R}|} \gamma_0 \gamma_5.
\end{equation}
Here $\boldsymbol{r}$ and $\boldsymbol{R}$ are positions of the electron and the nucleon, respectively, and $g_N^s$ is the scalar axion-nucleon coupling constant; $N$ can be either proton or neutron. 
As it was noted above, 
the significant difference between $V_{eN}$ and $V_{ee}$ is that $V_{ee}$ is the 
two-electron operator, whereas $V_{eN}$ and other often studied $T,P$-violating potentials depend on one electron. 

In the manner, analogous to Refs. \cite{Dmitriev:92,Maison:2021}, we introduce the molecular constant of the considered interaction:
\begin{equation}
    \Waxee(m_a) = \frac{1}{\Omega} \frac{1}{g_e^s g_e^p}
    \langle \Psi | 
    \mathop{{\sum}'}_{i,j=1}^{N_e} V_{ee}(\boldsymbol{r}_i, \boldsymbol{r}_j) | \Psi\rangle
\end{equation}
Here $\Psi$ is the electronic wavefunction of the molecule, $N_e$ is the number of electrons and $\Omega$ is the projection of the total electronic angular momentum $\textbf{J}_e$ on the molecular axis. For the YbOH molecule in the ground state $\Omega=1/2$. The summation is performed over all the electron pairs and the prime index means the terms with $i = j$ should be omitted. 

The $\Waxee$ constant cannot be measured, but it is required for the interpretation of experimental results in terms of the $g_e^s g_e^p$ product. The energy shift 
of the considered electronic state
caused by the interaction (\ref{potential}) can be expressed as
\begin{equation} \label{energyShift}
    \delta E = \Omega g_e^s g_e^p \Waxee(m_a)
\end{equation}

Therefore, it is possible to extract the $g_e^s g_e^p$ product from the measured value of $\delta E$ if the value of $\Waxee$ is known. \Eref{energyShift} is similar to the equation for the energy shift, caused by the $e$EDM, $T,P$-violating nucleus-electron contact scalar-pseudoscalar interaction or nucleon-electron axion-mediated interaction \cite{Dmitriev:92,Mosyagin:98,Skripnikov:16b,Skripnikov:14c,Skripnikov:17c,Fleig:17,Skripnikov:15a,denis2015theoretical, Maison:2021}:
\begin{equation} \label{eEDMcontribution}
    \delta E = \Omega d_e W_d,
\end{equation}
\begin{equation}\label{SPScontribution}
    \delta E = \Omega k_{T,P} W_{T,P},
\end{equation}
\begin{equation} \label{axioneNcontribution}
    \delta E = \Omega \bar{g}_N^s g_e^p \WaxeN(m_a).
\end{equation}
Here $d_e$ is the value of $e$EDM, $k_{T,P}$ is the dimensionless scalar-pseudoscalar nucleus-electron interaction constant, 
$\bar{g}_N^s$ is the scalar axion-nucleon coupling constant, averaged over all the nucleons in the heavy nucleus, 
$W_d$, $W_{T,P}$ and $\WaxeN$ are corresponding molecular constants. 

For QCD axions the upper limit for the axion mass was estimated to be $m_a \lesssim 1\  \textrm{meV}$~\cite{Peccei2008}. Under this condition, the interaction potential can be simplified using the approximation:
\begin{equation} \label{approx}
    \frac{e^{-m_a |\boldsymbol{r}_1 - \boldsymbol{r}_2}|}{|\boldsymbol{r}_1 - \boldsymbol{r}_2|} \simeq \frac{1}{|\boldsymbol{r}_1 - \boldsymbol{r}_2|} - m_a.
\end{equation}
Indeed, in order to use \Eref{approx}, the inequality $R_{\textrm{Yu}} \gg R_{\textrm{mol}}$ must be fulfilled, where $R_{\textrm{mol}}$ is the molecule size and $R_{\textrm{Yu}} = \hbar /(m_a c)$ is the characteristic range of the Yukawa-like interaction. The characteristic size of the YbOH molecule is 
3$\angstrom$ (see below), so the corresponding region for the axion mass is $m_a \ll 1\ \textrm{keV}$. 
For this case, the $\Waxee$ value is almost independent on $m_a$~\cite{Stadnik:2018}. Therefore, we present results obtained for $m_a = 1$ meV, but the same values are valid for lower masses $m_a$.

The difference of \Eref{energyShift} and Eqs. (\ref{eEDMcontribution}),(\ref{SPScontribution}) is that constants $W_d$, $W_{T,P}$ are determined solely by the electronic structure of the molecule, while $\Waxee$ is also a function of $m_a$. Therefore, generally the ALP mass requires an independent determination in this approach. As it was noted above, we are interested in the low-mass ALPs for which the $\Waxee$ constant is almost independent on $m_a$. Under this assumption the $g_e^s g_e^p$ product can be extracted from the measured energy shift using \Eref{energyShift} without a precise determination of $m_a$.

The electronic Hamiltonian in the Dirac-Coulomb-Gaunt approximation is given by
\begin{equation}
    H_{\textrm{el}} = 
     \Lambda^{(+)}
     \left[
    \sum\limits_{i=1}^{N_e} h(i) +   
    \mathop{{\sum}}_{i<j}^{N_e} 
    \left(
        \frac{1}{r_{ij}} - 
        \frac{\left( \boldsymbol{\alpha}_i, \boldsymbol{\alpha}_j\right)}{r_{ij}}
    \right)
    \right]
\Lambda^{(+)},
\end{equation}
where $h(i)$ stands for the one-electron Dirac Hamiltonian. The $i$ and $j$ indices run over
all electrons of the molecule. $r_{ij}$ is the distance between $i$-th and $j$-th electrons, $\boldsymbol{\alpha}_i$ is the vector of Dirac $\alpha$-matrices. $\Lambda^{(+)}$ are projectors on the positive-energy states. The first term in brackets is the Coulomb electron-electron interaction, and the second term is the magnetic two-electron interaction, also known as the Gaunt interaction. Its relative contribution to the molecular constants characterizing the $T,P$-violating effects in molecules is usually small \cite{Maison:20a, denis2020enhanced}. For this reason, below we exploit mainly the Dirac-Coulomb Hamiltonian and consider the Gaunt term as a correction.

The incorporation of the interaction (\ref{potential}) into relativistic electronic structure calculations implies the evaluation of the corresponding two-body integrals with molecular bispinors. Our calculations employed the expansions of the components of bispinors in Gaussian basis sets, and the integrals were first calculated in the basis of primitive Gaussian functions and then straightforwardly transformed to the bispinor representation. Approximation (\ref{approx}) allows one to replace the evaluation of the Yukawa-type integrals by the calculation of Coulomb-like and overlap ones which could be readily performed using the {\sc Libcint} library \cite{Valeev:libcint}.

High precision correlated electronic structure calculations were performed within the relativistic Fock space coupled cluster approach with single and double cluster amplitudes (FS-RCCSD) \cite{Visscher:01}. The ground state of the YbOH$^+$ molecular ion was considered as the closed-shell Fermi vacuum, and the $X\ ^2\Sigma_{1/2}$ ground state of the neutral YbOH molecule was obtained as the lowest solution in the $0h1p$ (one particle) Fock space sector. The active space comprised the two lowest Kramers pairs of molecular spinors. At the FS-RCCSD calculation stage all electrons were correlated. All the virtual spinors obtained at Dirac-Fock calculation stage were treated at the FS-RCCSD level. It should be emphasized that the inclusion of the high-energy virtual spinors is important for a thorough description of spin polarization and correlation effects for core electrons \cite{Skripnikov:17a,Skripnikov:15a}. All coupled cluster calculations were performed within the {\sc exp-t} program system \cite{Oleynichenko_EXPT, Oleynichenko:20}. Dirac-Hartree-Fock calculations of the YbOH$^+$ molecular spinors were carried out using the local version of the {\sc dirac15} software \cite{DIRAC15}. The linear geometry of YbOH with fixed values of internuclear distances $R(\textrm{Yb-O}) = 2.037 \angstrom$ and $R(\textrm{O-H}) = 0.951 \angstrom$ derived from spectroscopic experiments \cite{brutti2005mass, nakhate2019pure} was assumed in all calculations.

The Gaunt interaction contribution was calculated using the code, developed in Ref. \cite{Maison:2019} and extended to the case of molecular systems in Ref. \cite{Maison:20a}. The code for calculations of the $V_{ee}$ operator matrix elements (Eqs. (\ref{potential}) and (\ref{approx})) was developed in the present work. In order to calculate the $\Waxee$ constant value, we employed the finite field approach \cite{Monkhorst:77}, previously used to evaluate expectation values of one-electron operators.

\section{Results and discussion}
The values of the $\Waxee$ constant calculated at different levels of theory are summarized in \tref{ResultsTable}. To make sure the results obtained in the paper are convergent with respect to the basis set size, three basis sets of increased quality were used (see \tref{Basises}). Contribution of the Gaunt interaction was calculated within the basA basis set. This effect was considered simultaneously with correlation effects, i.e. beyond the Dirac-Fock-Gaunt approximation.

\begin{table}
    \caption{Notation and structure of the basis sets used.}
    \label{Basises}
    \begin{tabular}{lll}
    \hline
    \hline
         Basis set & Basis on  & Basis on  \\
         notation & Yb \cite{gomes:2010}$^*$ &  O and H \cite{Kendall:92,Dunning:89,de2001parallel}$^*$ \\
         \hline
           basA & AE2Z & aug-cc-pVDZ-DK \\ 
           & [24$s$,19$p$,13$d$,8$f$,2$g$] & [10$s$,5$p$,2$d$] and [5$s$,2$p$] \\
         basB & AE3Z & aug-cc-pVDZ-DK \\ 
          & [30$s$,24$p$,16$d$,11$f$,4$g$,2$h$] &
          [10$s$,5$p$,2$d$] and [5$s$,2$p$] \\
         basC & AE3Z & aug-cc-pVTZ-DK\\
          & 
          [30$s$,24$p$,16$d$,11$f$,4$g$,2$h$]
         & [11$s$,6$p$,3$d$,2$f$] and [6$s$,3$p$,2$d$] \\
    \hline
    \hline
    \end{tabular}
    $^*$All the basis sets were taken in the uncontracted form.
\end{table}

\begin{table}
    \centering
    \caption{The values of $\Waxee$ parameter within various approaches and basis sets.
    The results are presented in units of $m_e c /\hbar$.}
    \label{ResultsTable}
    \begin{tabular}{lccc}
        \hline
        \hline
         & basA & basB & basC \\
        \hline
         Dirac-Fock & $+1.22 \times 10^{-5}$  & $+1.22 \times 10^{-5}$  & $+1.22 \times 10^{-5}$  \\
         FS-RCCSD, DC & $+1.42 \times 10^{-5}$ & $+1.45 \times 10^{-5}$ & $+1.45 \times 10^{-5}$ \\
         FS-RCCSD, DCG & $+1.43 \times 10^{-5}$ & & \\ 
          \hline
          \hline
    \end{tabular}
\end{table}

As in the case of the electron-nucleon interaction \cite{Maison:2021}, the value of the parameter is weakly dependent on the basis set for low-mass axions. Comparison with results of Ref. \cite{Maison:2021} shows also that the ratio $\Waxee / \WaxeN \approx 0.43$ is close to $Z/A \approx 0.4$, where $Z=70$ is the charge of the Yb nucleus and $A=174$ is its nucleon number.
Such a ratio is explained in Ref. \cite{Stadnik:2018}.

The final result for $\Waxee$ is calculated as:
\begin{multline}
    \Waxee(\textrm{final}) = \Waxee(\textrm{basC}, \textrm{DC}) + \\
    +
    \left(
        \Waxee(\textrm{basA}, \textrm{DCG}) - \Waxee(\textrm{basA}, \textrm{DC})
    \right)=
    \\
    = 1.46 \times 10^{-5} \frac{m_e c}{\hbar}
\end{multline}
Here DC and DCG mean the Dirac-Coulomb and Dirac-Coulomb-Gaunt Hamiltonians, respectively. The term in brackets is the estimate of contribution of the Gaunt interaction effect. As one can see from \tref{ResultsTable}, the basis set dependence and Gaunt contribution are rather small (lower than 3\% and 1\%, respectively), so the main uncertainty is associated with neglected triple and higher-order excitations in the FS-RCCSD approach. It can be estimated to be lower than 10\% \cite{Maison:2019b,Maison:20a,Maison:2021}.

Current laboratory constraints on the $g_e^s g_e^p$ product can be derived from the ACME experiment on the ThO molecule \cite{ACME:18}.
According to the updated results of Ref.~\cite{Stadnik:2018}, it can be estimated as
$|g_e^s g_e^p|/(\hbar c) \lesssim 2.4 \times 10^{-19}$.
Substitution of this constraint and the final result for $\Waxee$ into \Eref{energyShift} gives the value of corresponding energy shift $\delta E \approx 220~\mu\textrm{Hz}$. 
This value 
is close to
the upper limit of energy shift measured using the ThO molecule~\cite{ACME:18}.
The sensitivity of the $T,P$-violation effects search experiment with YbOH is expected to be three orders of magnitude better \cite{kozyryev2017precision}. Thus, it can be expected that the updated constraints on $g_e^s g_e^p$ can be achieved in the YbOH experiment.

The estimated energy shift $\delta E$ induced in YbOH due to the considered interaction (\ref{potential}) can be compared with the shifts (\ref{eEDMcontribution})--(\ref{axioneNcontribution}) induced by other $T,P$-violating sources in this molecule and studied in previous works.
The values of these shifts and corresponding molecular constants for YbOH are summarized in \tref{ptOddShifts}. 
Note, that constraints on the $T,P$-violating parameters $d_e$, $k_{T,P}$, $\bar{g}_N^s g_e^p$ and $g_e^s g_e^p$ in Eqs.~(\ref{energyShift}) -- (\ref{axioneNcontribution}) used to estimate $\delta E$ for YbOH were extracted from the constraint on the $T,P$-violating energy shift obtained in the ThO experiment~\cite{ACME:18} and the values of the molecular constants for ThO corresponding to these $T,P$-violating sources calculated in Refs.~\cite{Skripnikov:16b,Fleig:16} for 
$d_e$ and $k_{T,P}$ and estimated in Ref.~\cite{Stadnik:2018} 
for $\bar{g}_N^s g_e^p$ and $g_e^s g_e^p$.
%
%

\begin{table}[h!]
    \centering
    \label{ptOddShifts}
    \caption{Energy shifts induced by various $T,P$-violating sources in the YbOH molecule for its ground electronic state. $X$ is the constraint on $T,P$-violating parameter corresponding to the  data from the ThO beam experiment~\cite{ACME:18} and $W_X$ is the corresponding molecular constant.}
    \begin{tabular}{lll}
        \hline
        \hline
        $T,P$-violating $\ $& $X$ and $W_X$ & $\delta E$, $\mu$Hz \\
        source & & \\
        \hline
        $e$EDM & $d_e \leq 1.1 \times 10^{-29} e \cdot \textrm{cm}$ &  60\\
               & $W_d = -47 $ GV/cm
               \cite{denis2019enhancement, gaul2020ab,prasannaa2019enhanced}&   \\
        SPS & $|k_{T,P}| \leq 1.9 \times 10^{-9}$ & 40\\
            & $W_{T,P} = -41 $ kHz \cite{gaul2020ab}& \\
        axion $e$N & $|\bar{g}_N^s g_e^p| \lesssim 1.0\times 10^{-19} \hbar c$ & \\
        ($m_a \ll 1$keV) & $\WaxeN = 3.36 \times 10^{-5} m_e c/\hbar \ $ & 200  \\
        & \cite{Maison:2021} & \\
        axion $ee$ & $|g_e^s g_e^p| \lesssim 2.4 \times 10^{-19} \hbar c $ &  220\\
        ($m_a \ll 1$keV) & $\Waxee = 1.46 \times 10^{-5} m_e c/\hbar$ & \\
        & (present work) & \\
        \hline
        \hline
    \end{tabular}
\end{table}

\begin{acknowledgments}
    Electronic structure calculations have been carried out using computing resources of the federal collective usage center Complex for Simulation and Data Processing for Mega-science Facilities at National Research Centre ``Kurchatov Institute'', http://ckp.nrcki.ru/.

    $~~~$Molecular coupled cluster electronic structure calculations have been supported by the Russian Science Foundation Grant No. 19-72-10019. Calculations of the $\Waxee$ matrix elements were supported by the foundation for the advancement of theoretical physics and mathematics ``BASIS'' grant according to Projects No. 20-1-5-76-1 and No. 18-1-3-55-1. Calculation of the Gaunt contribution has been supported by Russian Foundation for Basic Research Grant No. 20-32-70177.
\end{acknowledgments}

\bibliographystyle{apsrev}

\end{document}